\documentclass[prd,aps,amsmath,amssymb,showpacs]{revtex4}
\usepackage{graphicx}
\usepackage{epsfig,rotate,latexsym}
\begin{document}
\title{Probing the anisotropic expansion history of the universe
with cosmic microwave background}
\author{Ranjita K. Mohapatra}
\email {ranjita@iopb.res.in}
\author{P. S. Saumia}
\email {saumia@iopb.res.in}
\author{Ajit M. Srivastava}
\email{ajit@iopb.res.in}
\affiliation{Institute of Physics, Sachivalaya Marg, 
Bhubaneswar 751005, India}
%
%

\begin{abstract}

We propose a simple technique to detect any anisotropic expansion stage 
in the history of the universe starting from the inflationary stage to 
the surface of last scattering from the CMBR data. We use the property that 
any anisotropic expansion in the universe would deform the shapes of the 
primordial  density perturbations and this deformation can be detected 
in a shape analysis of superhorizon fluctuations in 
CMBR. Using this analysis we obtain the constraint on any previous 
anisotropic expansion of the universe to be less than about 35 \%. 
\end{abstract}

\maketitle

\section{INTRODUCTION}

Observations show that the present universe is homogeneous and
isotropic on large scales. One of the most important observational 
evidences for the homogeneity and isotropy of the universe is the 
highly smooth and uniform cosmic microwave background radiation (CMBR).
The distribution of matter in the universe is also generally consistent 
with a homogeneous and isotropic universe at present.  However, this does 
not rule out the possibility of an anisotropic expansion of the universe 
during some early stage (say, during inflation) such that the expansion 
becomes isotropic later on. For example, the universe may have expanded 
anisotropically during the early stages of the inflation. If the 
expansion becomes isotropic subsequently (which is most likely to happen 
due to the no-hair theorem \cite{wald}), then it is not
obvious whether any signatures of such an early, transient, stage of
anisotropic expansion would be present in the present day universe.
We ask the question whether one can put constraints on such an early, 
transient, stage of anisotropic expansion using CMBR data in, as much 
as possible, model independent way (e.g. not even assuming inflationary stage).
We propose a technique for this which is based on analyzing the
shapes of  superhorizon sized fluctuations in
CMBR at the surface of last scattering, i.e. 
patches of angular sizes larger than about 1 degree at the present stage.
Due to being superhorizon until the last scattering stage, the
shapes of these patches will undergo simple scaling in different
directions, and hence will retain the memory of any anisotropic 
expansion stage in the entire history of the universe. Though even the 
subhorizon fluctuations in CMBR may retain
this shape information due to linear evolution. However, for 
subhorizon fluctuations one cannot rule out various causal effects
(such as the presence of magnetic fields) which may distort the
shapes. Statistically such a distortion (from causal effects)
may not be significant, however, it seems simpler to restrict
to superhorizon fluctuations for the initial analysis.

Our technique is very general and does not depend on any specific models
for describing the early universe stages. For example, even in the absence of
any inflationary stage, the shapes of the density fluctuations patches at
superhorizon scales will contain the memory of any anisotropic expansion
stage. However, for the sake of definiteness, and with the general
confidence in an inflationary paradigm, we will continue to refer to
an early inflationary stage of the universe  which is responsible for
the generation of density fluctuations in the universe. Note also, though
we talk about detecting anisotropic expansion, any other anisotropy
which leaves its imprints in the universe in terms of shape deformations
(for example if initial density fluctuations themselves were anisotropic)
will also be detected by our technique. However, it is clearly important
to know if one can distinguish between the anisotropy originating
entirely from the initial conditions, and the one arising from 
anisotropic universe expansion. For example, in the inflationary
paradigm, our entire observed universe has originated from a single
causal patch. It seems entirely possible that initial fluctuations
may be anisotropic due to the presence of some background field.
Once these fluctuations go out of the horizon, they will then remain
anisotropic, and will eventually manifest in the anisotropic shapes
of patches of CMBR. (Though the effects of such a background field will
diminish rapidly, hence fluctuations produced later on will be
mostly isotropic.) One would like to distinguish such a situation of
initial anisotropic fluctuations from the case when fluctuations
are produced isotropically but undergo shape deformations due
to anisotropic expansion of the universe. (Even here, eventually the
universe expansion becomes isotropic \cite{wald}, hence any fluctuations 
produced after that will remain isotropic.) 

 We will discuss two different ways of identifying shape deformations
of temperature  fluctuation patches. We generate patches by considering
some threshold value for overdensity (or underdensity) of CMBR 
fluctuations in a given region on the surface of last scattering.
This will generate the excursion sets and we will get a picture of
that region consisting of these excursion sets. We then study shape 
deformation of these excursion sets (patches) in this region using 
either the Fourier transform, or we detect it directly in the
spatial patch. Both these methods can detect the shape deformations.
However, direct shape analysis with spatial region has specific
advantages which we will discuss later on. 
We have also analyzed simulated density fluctuations of specific
geometrical shapes, such as spheres, ellipsoids. This helps us
in analyzing the strengths of these two techniques in differentiating
various scenarios, such as anisotropies arising from initial
conditions, or anisotropies arising from the anisotropic expansion etc.
We will show that this simple shape analysis is able to 
answer an important question in an almost model independent manner. 
That is whether the universe ever expanded anisotropically almost from 
the beginning of inflation near $t \simeq 10^{-35}$ sec. all the way up 
to the stage of last scattering when the universe was 300,000 years old.

 The paper is organized in the following manner. In section II, we
will briefly discuss some other techniques which have been used in
the literature for studying the anisotropy of the universe. Section III
discusses the method for generating the excursion sets for the shape
analysis. Section IV discusses the shape anisotropy analysis using
Fourier transforms. In section V we describe our method for analyzing
shape anisotropy directly in the spatial region. Section VI discusses
applications of these techniques to detect anisotropic expansion by
artificially stretching the CMBR data in one direction. Using different
stretching factors we constrain the anisotropic expansion factor 
of the universe. Section VII discusses the analysis with simulated
fluctuations of definite geometric shapes and sizes. Different cases
are analyzed to distinguish different scenarios leading to anisotropic
shapes of fluctuations in the universe. Conclusions are presented 
in section VIII.

\section{OTHER PROBES OF THE ANISOTROPY OF THE UNIVERSE}

The density perturbations resulting from the general inflationary
picture (consequently temperature fluctuations in CMBR) are expected to be
statistically homogeneous and isotropic Gaussian random fluctuations around
a uniform background. Many people have discussed departures from 
the isotropy.  Non-Gaussianity is an important probe of various
inflationary models as well as contributions from other sources
of density fluctuations, and is being intensively investigated. 
Regarding the isotropy of the universe, recently there have 
been claims of evidences of statistical anisotropy in CMBR 
\cite{cmbanstrpy,aoe} on large angular scales. These evidences are more 
of the nature of identifying a special direction (axis) in the CMBR 
sky \cite{aoe} which shows excess power in specific $m$ modes in the
multipole expansion of  temperature fluctuations. Various suggested
explanations for this statistical anisotropy include nontrivial topology of
the universe \cite{tplg} and anisotropic expansion. As we mentioned, the 
anisotropy discussed in these works corresponds to large angular scales, 
e.g. referring to the alignment of quadrupole and octupole moments in 
the CMBR sky.  In contrast, our analysis method probes anisotropy at 
relatively shorter angular scales from $1^\circ$ to about $20^\circ$ 
(as we will explain below).

There are other techniques, e.g.  Minkowski functionals \cite{mnkwsk} which 
have been used for analyzing CMBR fluctuations (e.g.  non-Gaussianity 
\cite{mnkcmb}) and morphological and topological properties of matter 
distribution in the universe \cite{mnkstr}. These techniques are very useful 
in providing structural properties of density fluctuations. For example,
{\it shapefinders} can yield statistics of filamentary or pancake
structures. These techniques have also been used in diverse areas, e.g. for
investigating geometry and topology  of the excursion sets (with a given
threshold) for the magnetic field of the Sun \cite{mnksun}.
Such techniques applied to CMBR excursion sets will also
lead to useful information about the nature of density fluctuations,
for example Gaussianity, geometry and topology of excursion sets, etc.  
However, an overall shape deformation
of fluctuations arising from anisotropic universe expansion is an
entirely different issue and Minkowski functionals have not been used
to analyze this issue. For example, a randomly oriented collection
of filamentary structures (or pancake structures) will be detected
using shapefinders, but it does not imply anisotropic expansion of
the universe which will require an overall alignment of the orientation
of such structures. Our purpose is to detect such an overall alignment
of (statistically) non-spherical structures.  
Similar situation is with other techniques which have been used in
the literature to study topological features of CMBR fluctuations.
For example, the genus statistics of excursion sets of CMBR fluctuations
has been used to detect non-Gaussianity \cite{praba}. However, here also
any possibility of an overall orientation is not investigated. In this 
context we mention that there has been study of the ellipticity of 
anisotropy spots of CMBR, and the orientation of such ellipticity 
has also been discussed \cite{ellps}. This study was motivated by
the effects of geodesic mixing, though it has similarities with 
our technique in the sense of attempting to detect elliptic shapes and
orientations of such shapes. However, a systematic study of general 
shape deformations and orientation resulting from an anisotropic 
expansion the universe during entire history of the evolution of
density fluctuations in the universe has not been carried out.   

 The possibility of anisotropic expansion of the universe has been
discussed in the literature in different contexts. There are models which 
describe anisotropic cosmologies \cite{bianchi} and these have been
investigated from the point of view of constraining the anisotropy
parameters of the models from CMBR data \cite{bianchicmbr}. The possibility 
of anisotropic expansion due to strings, walls and magnetic fields has also 
been discussed \cite{ansinf}. It has also been discussed that with some
coupling between gauge fields and the inflaton, the anisotropy may persist
through the inflation  (avoiding the no-hair theorem \cite{wald}) and 
resulting power spectrum of fluctuations
has been calculated \cite{aniscont}.  We will
not be discussing any specific model here for anisotropic expansion of
the universe. We just assume that for whatever reason, if ever there
was an anisotropic expansion of the universe in the history of the universe,
from the beginning of the origin of density fluctuations (relevant to the
observational scales at present) all the way up to the stage of last 
scattering, then we ask how one can probe it using shape analysis of 
patches (excursion sets) of CMBR fluctuations.

\section{GENERATING EXCURSION SETS FOR SHAPE ANALYSIS}

 The main idea underlying our analysis is the following. If the density 
perturbations generated initially (say during inflation)  are 
statistically isotropic, then they will continue to have an average 
spherical shape (statistically) if the expansion of the universe was 
always isotropic. But if the universe ever went through an anisotropic 
expansion, then these perturbations would get deformed in a specific
direction \cite{ansinf}. For example, if the universe ever expanded 
differently in one direction from the other two then the {\it average} 
shape of these perturbations will become ellipsoidal, with the axes of
the (statistical) ellipsoids all aligned. This average deformation will 
survive the later isotropic expansion at least till the time the size of 
the perturbation remains superhorizon. (Assuming that these small
superhorizon fluctuations evolve linearly and undergo simple scaling
with the expansion.) This means that any anisotropic expansion 
from the beginning of inflation to the surface of last
scattering will leave their signature in the shapes of superhorizon 
perturbations in the CMBR. Fig.1 shows this basic physical picture where
initially spherical fluctuations are shown to evolve with isotropic
expansion of the universe and with a transient anisotropic expansion
stage. The fluctuations remain spherical if the universe expansion
was always isotropic. However, they deform to ellipsoidal shapes 
during an anisotropic expansion period (which, for simplicity, is taken
here as faster expansion along the major axis of the ellipses shown).
Subsequently the expansion of the universe becomes isotropic. However,
the fluctuations retain their ellipsoidal shapes, with same geometry,
as well as orientation. Note here we are referring to superhorizon
fluctuations (of small amplitude), hence they are assumed to evolve linearly 
with the scale factor. This shape deformation information may also remain 
in the sub-horizon fluctuations due to linear evolution of the fluctuations.
However, there may be causal effects, e.g. due to the presence of
background fields, which may affect the shapes of these fluctuations.
Though such causal effects will still remain uncorrelated over superhorizon
scales. Hence they should not affect the statistical isotropy of the 
fluctuations. However, it seems simpler to rely on superhorizon 
fluctuations for the detection of anisotropic universe expansion.
With high resolution data from Planck it will be important to consider
shapes of sub-horizon fluctuations, especially as it will contain
information about the dynamics of various causal factors present during
the acoustic oscillations.

\begin{figure*}[!hpt]
\begin{center}
\leavevmode
\epsfysize=7truecm \vbox{\epsfbox{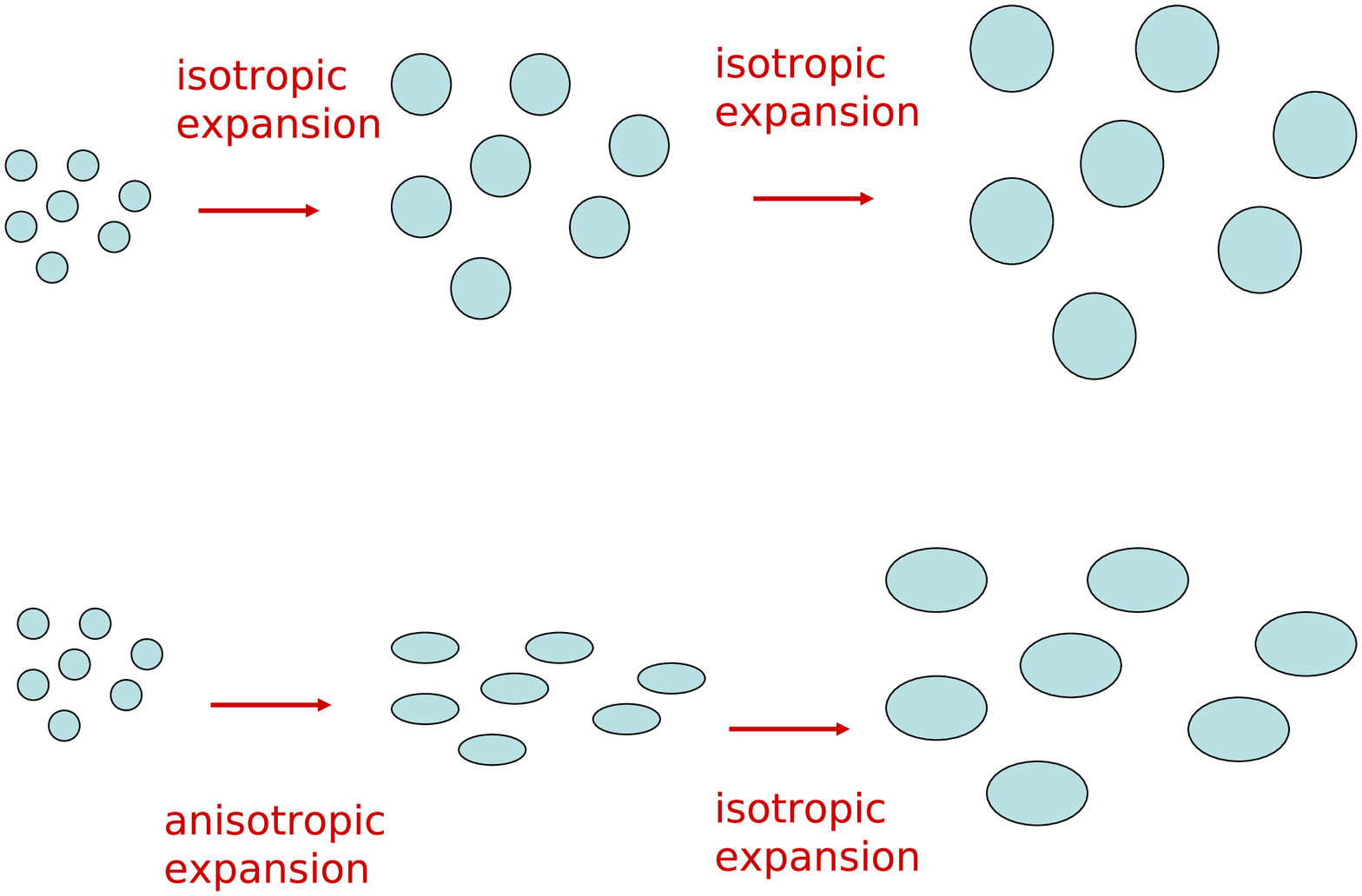}}
\end{center}
\caption{}{Basic physical picture of our model. Top part shows the evolution
of initially spherical fluctuations during isotropic universe expansion.
Fluctuations remain spherical. Bottom part shows the effects of a transient
anisotropic expansion stage of the universe on the initially spherical
fluctuations. Fluctuations become ellipsoidal with the same orientation,
and these features are retained subsequently.}
\label{Fig1}
\end{figure*}

For an anisotropic inflation which later tends towards isotropy, deformation 
will be more for large size patches, those which exited the horizon during 
early stages of the inflation. The shape deformations will tend to zero for 
the small size patches, those which exit the horizon once the universe
expansion becomes (almost) isotropic. Thus, a  careful analysis of the 
shapes of patches of different sizes may be able to show this 
pattern in the CMBR sky of an anisotropic universe tending to an isotropic 
one (unless anisotropic expansion stage happened too early, followed by
a very large duration of isotropic inflationary expansion). Any 
anisotropic expansion after the inflation (more precisely, after the
generation of fluctuations) can always be detected by our 
method, as in this case patches of all superhorizon sizes will show 
similar deformation.

We look for these shape deformations in the over/under density 
patches with sizes greater than about $1^\circ$ in the CMBR sky by measuring 
their extents along and transverse to a given axis. We consider a 
relatively thin strip along the equator of the CMBR sky (so that it is 
reasonably flat for  projection on a plane), and analyze the extents of 
various patches along the latitudes (called as the X direction) and along 
longitudes (called as the Y direction).  Thus our analysis is not very 
accurate for very large scale anisotropies for which one will need 
to analyze anisotropies in patches covering large parts of the sphere.
If the universe ever expanded faster or slower along the polar axis 
compared to its expansion in the equatorial plane, then the shape 
deformations of density (temperature) fluctuation patches along this thin 
equatorial belt will be most sensitive to it. For example, the stretching
of small patches at the equator will be simply proportional to
the ratio of expansion factors along z axis that along the transverse 
equitorial plane. This is what we will assume in the following
when setting constraints on the expansion factor anisotropy of the
universe. Note that for patches of large angular scales this 
proportionality will not hold. We repeat this analysis for various 
orientations of the north pole in the CMBR sky, thereby probing 
anisotropic expansion in any possible direction. 

\begin{figure*}[!hpt]
\begin{center}
\leavevmode
\epsfysize=10truecm \vbox{\epsfbox{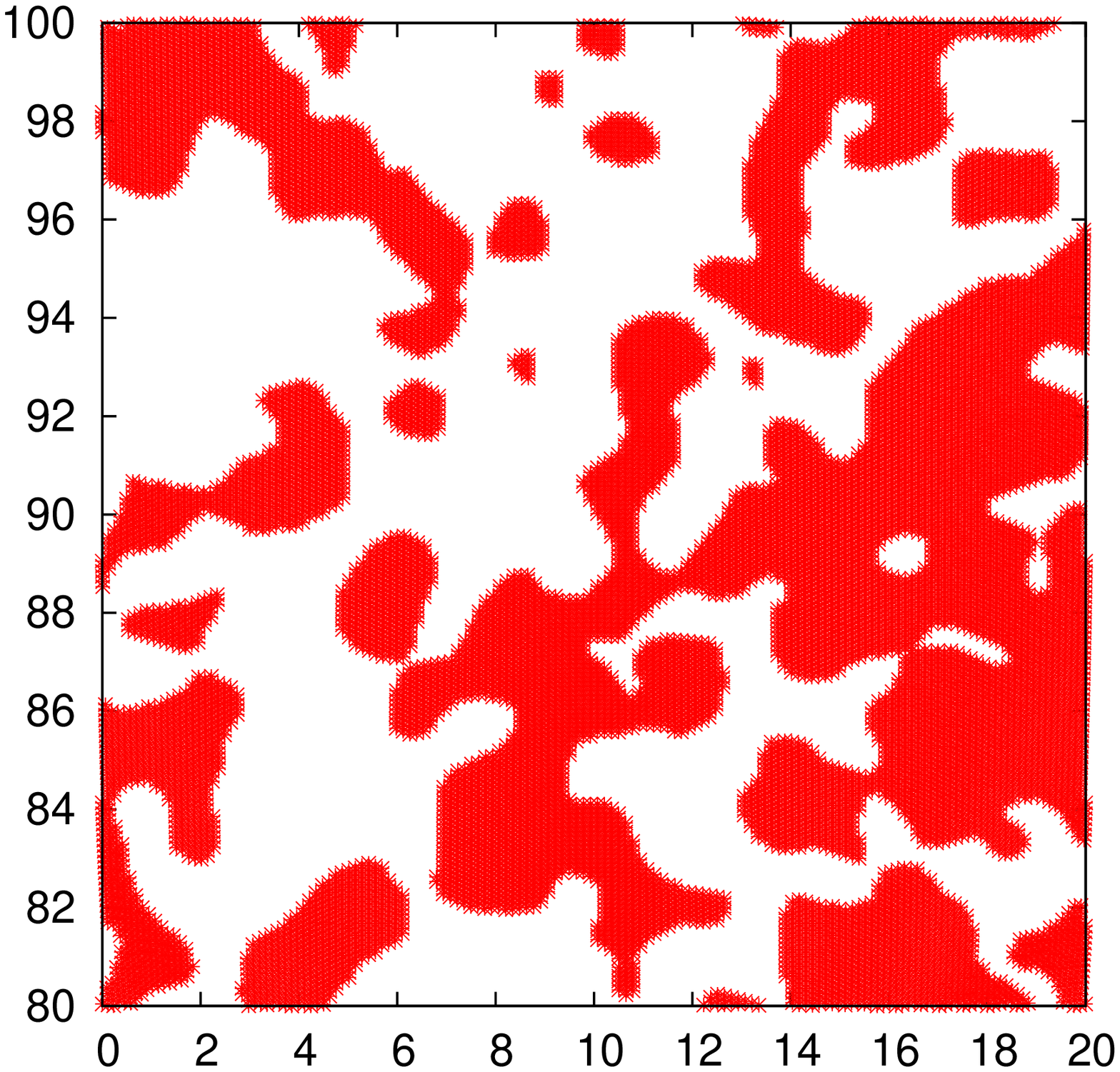}}
\end{center}
\caption{}{All the angles are in degrees. In all the figures, the  X axis
corresponds to $\phi \sin\theta$ and the Y axis shows $\theta$. This figure
shows the overdensity patches (excursion sets) in a $20^\circ \times 
20^\circ$ region in the equatorial belt of CMBR sky (with a suitable 
threshold value).}
\label{Fig2}
\end{figure*}

 We use WMAP 7-Year Data (ILC map) \cite{wmap7} for the analysis. However,
we face a problem in applying our technique by using a thin strip along
the equator. This is because the equatorial region with WMAP data is
the most contaminated one with foreground noise. We avoid this problem
by rotating the z axis (about the x axis) by 30$^\circ$ and by 
70$^\circ$, (due to special arrangements of pixels in HEALPix, 
some orientations are not convenient, e.g. 90$^\circ$). Then we
carry out our analysis by considering a strip
along the (rotated) equator. There will still be some contamination from
the intersection points with the original equator, but hopefully
its contribution will be relatively small. We choose a strip of width
$\pm 10^\circ$ along this rotated equator, and choose CMBR fluctuations 
above/below a particular value thereby forming the excursion sets. 
Fig.2 shows a small $20^\circ \times 20^\circ$ region in this equatorial 
belt.  Filled patches seen in Fig.2 correspond to temperature anisotropies
of magnitude (0.02 - 1)$\times (\Delta T)_{max}$ where $(\Delta T)_{max}$
is the maximum magnitude of CMBR temperature anisotropy in this patch. 
As one can see, the patches appear randomly shaped and sized. This picture
is projected on a plane to calculate the X and Y extents of the filled
patches (with the X extent being taken as $\phi \sin \theta$ and the Y 
extent is taken as $\theta$). $\theta$ and $\phi$ are the polar and 
azimuthal angles respectively. Thus, though we denote X and Y in terms of 
$\phi\sin\theta$ and $\theta$ in degrees, these
actually represent length scales with suitable multiplication of radius
of a sphere representing last scattering surface.

  We see that actual fluctuations in CMBR in Fig.2 do not look anywhere as 
simple as the geometrical fluctuations in Fig.1, and one cannot 
straightforwardly look for the shape deformations, and any overall orientation 
of such deformations. Even though fluctuation patches in Fig.2 appear
of arbitrary shapes, statistically averaged shape of the fluctuations 
will be spherical (isotropic), and main idea of our approach is
to detect any deformations in this {\it statistically averaged
shape} of the fluctuation patches. Thus, isotropic expansion will stretch 
the fluctuations symmetrically, leading to the statistically averaged
shape remaining spherical, while any anisotropic expansion stage 
stretches the fluctuations asymmetrically.  The statistically averaged 
shape will become ellipsoidal if the expansion is asymmetric along one 
direction.

\section{FOURIER TRANSFORM TECHNIQUE}

 Detection of shape anisotropies using Fourier transforms can be done
in different ways depending on what criterion one adopts to characterize
the anisotropy \cite{ft,ftfiber,ftmtlrg}. The techniques of refs.
\cite{ftfiber,ftmtlrg} are somewhat similar and here we will follow 
the approach used in \cite{ftmtlrg} for analyzing anisotropic 
deformations in metallography. In ref. \cite{ftmtlrg},
a digitized image of the material was used and 2D Fourier transform of this
image was calculated and subsequently  thresholded to levels 0 and 1. 
Anisotropy in the Fourier space was then determined by 
the ratio of the widths of the histograms in the two directions.
 
 Fig.3a shows the 2D Fourier transform of the picture in Fig.2 where
we have used a suitable threshold to convert the values of the Fourier
transform to 0 and 1. Fig.3b shows the plots of the histograms in the
two directions $(k_x,k_y)$ corresponding to Fig.3a. Solid curve and the 
dashed curves are the best fits to the histograms, which are Gaussian 
curves. The two plots completely overlap. The ratio of the widths of the 
Gaussians in $k_x$ and $k_y$ directions is

\begin{equation}
{\sigma_{k_x} \over \sigma_{k_y}} = 1.03 \pm .02
\end{equation}

showing the statistical isotropy of the excursion sets in Fig.2.
We mention here that the shapes of these two curves depends on the
threshold values used for converting the Fourier Transform to
levels 0 and 1. In our analysis (for all the figures) we have tried
to fit (un-normalized) Guassians as their widths and the normalizations
provide a quantitative characterization of the anisotropy. However for
certain cases (e.g. in Figs.10, and 11 below) it is not possible to
fit Gaussians. More appropriate curves  are of the shape of
Woods Saxon potential and that is what we use. in such cases, we use
suitable threshold values which maximize the difference between
the two histograms, while the histograms remain reasonably smooth
curves. 

\begin{figure*}[!hpt]
\begin{center}
\leavevmode
\epsfysize=10truecm \vbox{\epsfbox{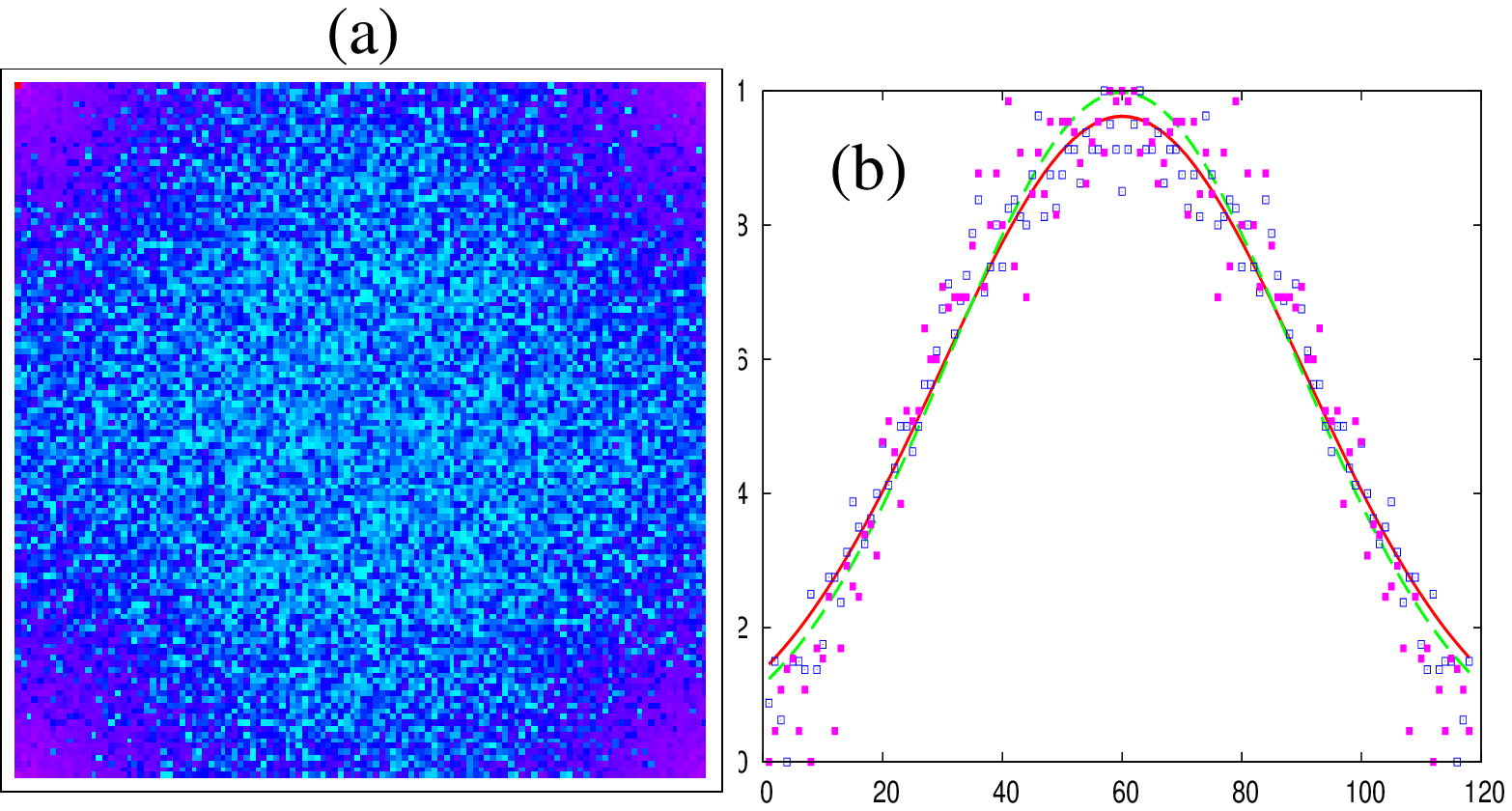}}
\end{center}
\caption{}{(a) shows the 2D Fourier transform of the picture in Fig.2
with an appropriate threshold to levels 0 and 1, with the horizontal and
the vertical axes representing $k_x$ and $k_y$ (this picture is shown only
to qualitatively see the isotropy of the scatter plot, hence we we do not 
show axis markings here). (b) shows the 
histograms in the two directions for (a). Solid and dashed curves (which
completely overlap) are best fits to these histograms.}
\label{Fig3}
\end{figure*}

\section{SHAPE ANALYSIS IN THE PHYSICAL SPACE}

 We now describe our technique for detecting shape deformations directly
in the physical space. Clearly, what one wants to know is the average
width of the fluctuations in X direction and compare it with the
average width in the Y direction. That will work well if fluctuations
were of specific geometrical shapes and sizes and not  for fluctuations
as in Fig.2. What one wants to know here is the distribution of
widths in the X direction and compare it with the distribution of widths
in the Y direction. For this we proceed as follows. We divide the entire 
$20^\circ$ wide equatorial belt (with $\theta$ within $\pm 10^\circ$
about the equator, and $\phi$ ranging from 0 to $360^\circ$) into thin 
slices (varying from 0.02$^\circ$ - 0.1$^\circ$) in X and Y directions 
(to increase statistics). Using these slices, we determine the X and Y 
extents of various filled  patches. This procedure is shown in Fig.4a for
the $20^\circ \times 20^\circ$ region of Fig.2. (Note that X with 
sin($\theta$) factor, and 
Y widths, represent distances on the sphere. However, we keep referring to 
these in degrees.) We then plot the frequency distributions (histograms) 
of X and Y widths of the intersections of all the patches with these slices
in this entire equatorial belt. For the isotropic case, we expect the X and Y 
histograms to almost overlap. Any relative shift, or difference, between the 
X and the Y histogram will imply the presence of an anisotropy, such 
as an anisotropic expansion. We mention that this fine slicing of the 
equatorial strip is done to simplify the shape analysis. 
The main idea is to get statistical information
about X and Y widths of these random patches. The information on the nature
of expansion will be contained in the distributions of each X and Y slice
of the equatorial strip and collection of data from all the slices will
therefore provide us with a good statistics.

 Fig.4a shows these slices in the X and Y directions for the picture 
in Fig.2. Fig. 4b shows the frequency distributions (histograms) 
of the widths of the intersections of the slices with the filled 
patches (excursion sets) in X and Y directions for the entire equitorial
belt. The solid curve shows the 
histogram of X slices of filled patches and the dashed curve shows the 
histogram for the Y slices. The horizontal axis corresponds to the widths 
of the slices in degrees (for X slices, it represents width using
$\phi \sin\theta$), with histogram bin having width of $1^\circ$.
The vertical axis gives the frequency $N$ of the occurrence
of the respective widths in all the slicings (X or Y respectively) of 
excursion sets in the equatorial strip. The error bars denote the statistical
uncertainty of $\sqrt{N}$ for the frequency $N$ in each bin. 
We can see that the two histograms, corresponding to X and Y extents
of the patches, are almost overlapping. This confirms the statistical
isotropy of the fluctuations in Fig.2 (calculated here for the entire
equatorial belt). This is consistent with the isotropy seen in Fig.3 
using the Fourier transform of the $20^\circ \times 20^\circ$ region 
of Fig.2. (We mention here that the peak at about $1^\circ$ in Fig.4b
presumably corresponds to the smoothing of WMAP data below this scale.)

\begin{figure*}[!hpt]
\begin{center}
\leavevmode
\epsfysize=10truecm \vbox{\epsfbox{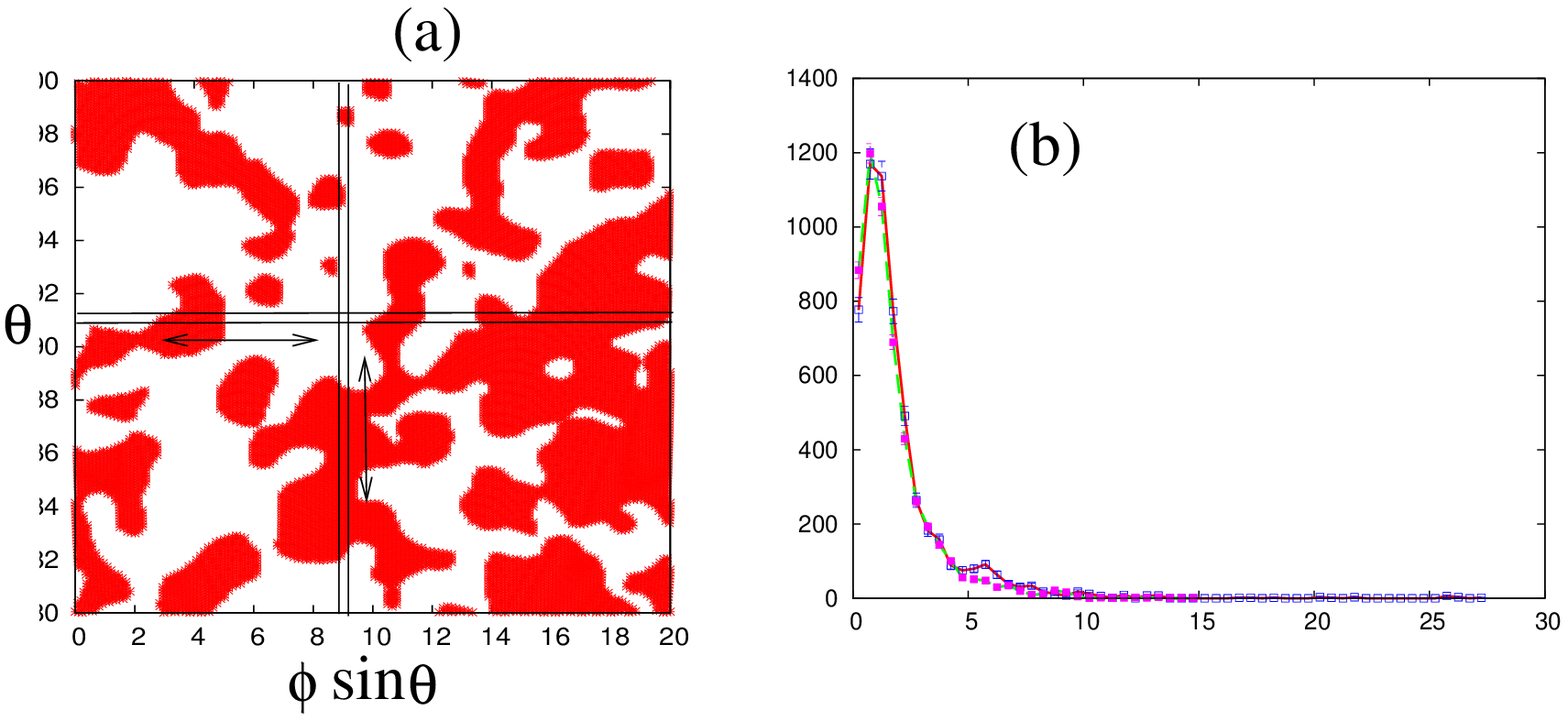}}
\end{center}
\caption{}{(a) Slicing used along X and Y axes for the $20^\circ \times
20^\circ$ region in Fig.2.
Widths of filled patches (excursion sets) intersected by the slices
are determined and histograms are calculated for the distribution of 
these widths in X and Y directions. (b) shows plots of these histograms
of the widths of filled patches obtained using the entire equitorial belt
with a width of 20$^\circ$ (along longitude). X axis represents widths of 
patches. Bin width is taken as  1$^\circ$. Smooth curves join 
the points which are marked with corresponding ($\sqrt{N}$) error bars. 
Distribution for width along X ($\phi\sin \theta$) is shown by solid 
plot and distribution for width along Y ($\theta$) is shown by the 
dashed line.}
\label{Fig4}
\end{figure*}

\section{CONSTRAINING ANISOTROPIC EXPANSION}

To determine the level of isotropy which is implied by the overlap of 
the two histograms in Fig.3 and in Fig.4, we introduce artificial stretching 
in the CMBR patches in the following manner and repeat the above analysis for 
this {\it stretched} data.
To {\it simulate} stretching by a factor $\alpha$  we simply multiply
the Y coordinate (i.e. $\theta$) for each point in the equatorial
strip used above by a factor $\alpha \times {|y| \over 10^\circ}$. Here $y$ 
denotes the value of $\theta$ in degrees measured with respect to the 
equator, with $-10^\circ \le y \le 10^\circ$ 
for the above equatorial strip. This stretching represents an anisotropic 
expansion of the universe along the polar axis compared to the expansion 
in the equatorial  plane by a factor $\alpha$.
Note again that this simple scaling expression works approximately fine for 
a relatively thin strip along the equator. For strips having larger widths
along the longitudes, the y coordinates of different patches will
be scaled by a more complicated factor.

\begin{figure*}[!hpt]
\begin{center}
\leavevmode
\includegraphics[totalheight=0.5\textheight, angle=-90]{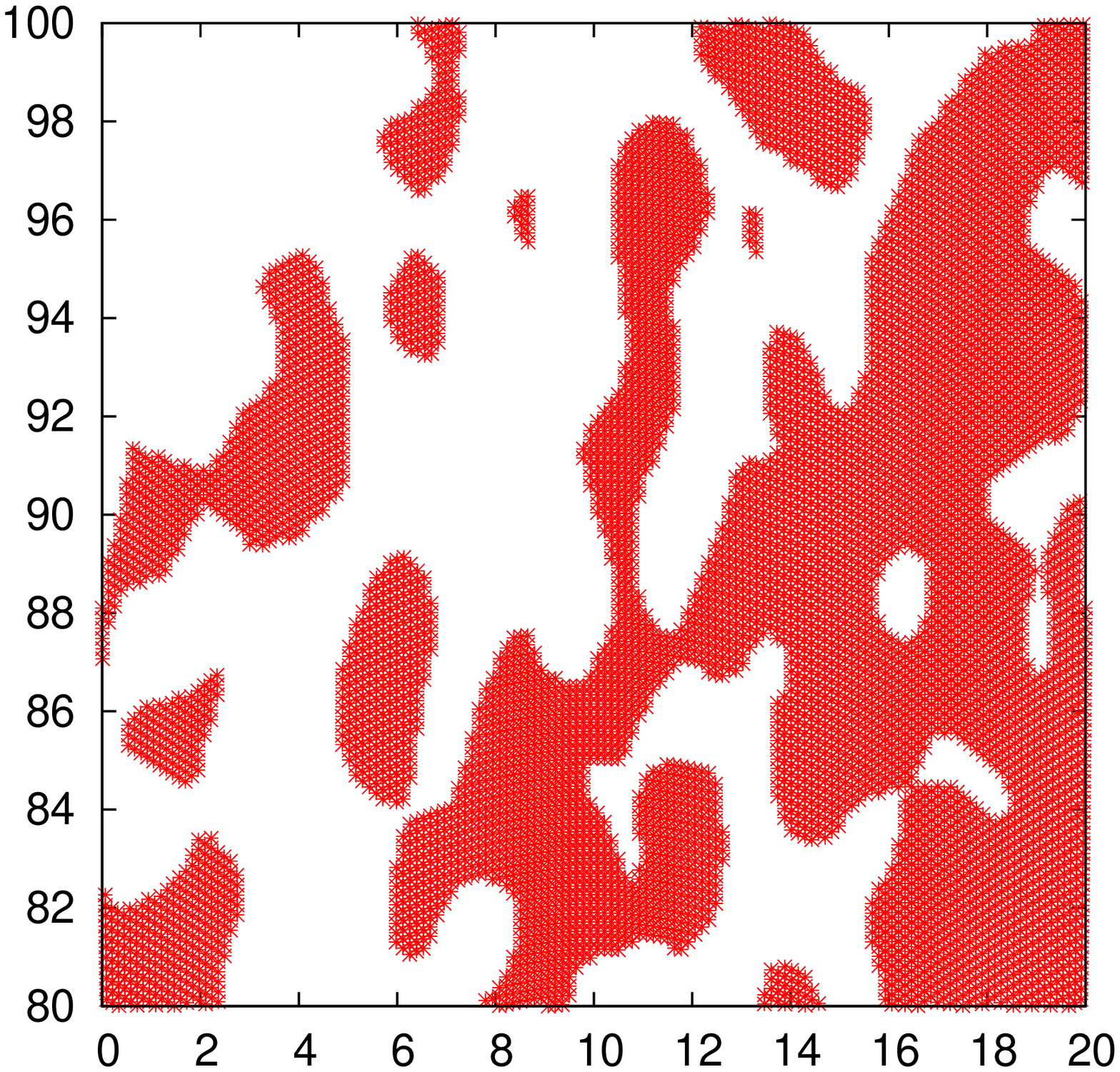}
\end{center}
\caption{}{This figure shows the overdensity patches (excursion sets) 
in a $20^\circ \times 20^\circ$ region in the equatorial belt of CMBR 
sky after being stretched by factor 2 along the z axis.}
\label{Fig5}
\end{figure*}

  Fig. 5 shows this artificially stretched patch corresponding
to the patch shown in Fig.2. The stretching factor is $\alpha = 2$
for Fig.5. For the consistency of the analysis we take the $20^\circ$ wide 
entire equatorial strip from this stretched data set (the stretched strip
now spans $40^\circ$ width along the longitude for $\alpha = 2$). We first
repeat the analysis of section IV by calculating the 2D Fourier
transform of the picture in Fig.5, with suitable threshold for levels 0 
and 1 and then calculating histograms of these values in the two 
directions. Similarly the analysis of section V is repeated for
Fig.5 by slicing the entire (stretched) equatorial strip in X and
Y (i.e. $\phi\sin \theta$ and $\theta$) directions and determine 
corresponding histograms for widths of the filled patches. 
The results are shown in Fig.6.  Fig.6a shows the Fourier transform
of the picture in Fig.5 and Fig.6b shows the histograms calculated
for this 2D Fourier transform as in Fig.3. We see that these histograms 
do not overlap showing the anisotropy arising from the stretching in Fig.5. 
The ratio of the widths of the Gaussians in $k_x$ and $k_y$ directions is
${\sigma_x \over \sigma_y} = 1.04 \pm .04$. Anisotropy of Fig.5 is reflected
here in the fact that the normalizations of the two Gaussians are
different. Ratio of these two normalizations is 0.85 $\pm$ .01 .

Similarly, Fig.6c shows the histograms resulting from the analysis
of section V applied to the stretched equitorial belt. 
Again, the solid and dashed curves correspond 
to the X and Y histograms respectively. We see here also that the two 
histograms do not overlap and the difference is significant.  The X 
(i.e. $\phi\sin\theta$) 
histogram fall above the Y (i.e. $\theta$) histogram in the first few bins 
which is due to the fact that we have stretched all the patches
thereby reducing the number of patches with small Y width. But for larger
widths we have the Y histogram crossing and remaining above the X histogram
which implies that we have more patches with larger width in Y slices.
Note incidentally that the  peak in the dashed curve ($\theta$ histogram)
has shifted to larger widths by about a factor of 2 which is the factor
of stretching of patches for Fig.5.

We mention here that we are not trying to deduce a number similar to the 
widths (or normalizations) of the histograms in Fig.3b, for some suitable 
widths of the curves in Fig.6c which can be compared for the X and Y 
distributions leading to a simple quantitative characterization of the 
anisotropy. This is because the precise nature of difference expected 
between the two histograms in Fig.6c is not easy to relate to the
stretching factor $\alpha$ as it may also depend on the geometrical
shapes of different random patches. For example, in the plots in Fig.6c,
there are significant changes in the two distributions
at larger angular scales, say, 10$^\circ$ - 20$^\circ$. However, these are
dominated by fluctuations. Also, as mentioned above, at this stage our
technique is not accurate for large scales. As we will see
in later sections where simulated fluctuations of various geometrical
shapes are analyzed, the changes in the $X$ and $Y$ histograms for pictures 
similar to  Fig.5 are of very different nature. Unless one understands
how the entire curve depends on the geometry, orientation, and distribution
of the patches, it does not seem useful to device any parametrization
of these curves.  For us, the significant point is that a stretching  
factor of $\alpha = 2$ produces X and Y histograms which are clearly 
distinct, beyond the error bars. A comparison
with Fig.4b for the actual CMBR data implies that
in the entire history of the universe, any anisotropy in its
expansion along the z axis is bounded by $\alpha < 2$.
In this sense of comparing the two distributions, our 
technique remains semi-qualitative.    

\begin{figure*}[!hpt]
\begin{center}
\leavevmode
\epsfysize=5.5truecm \vbox{\epsfbox{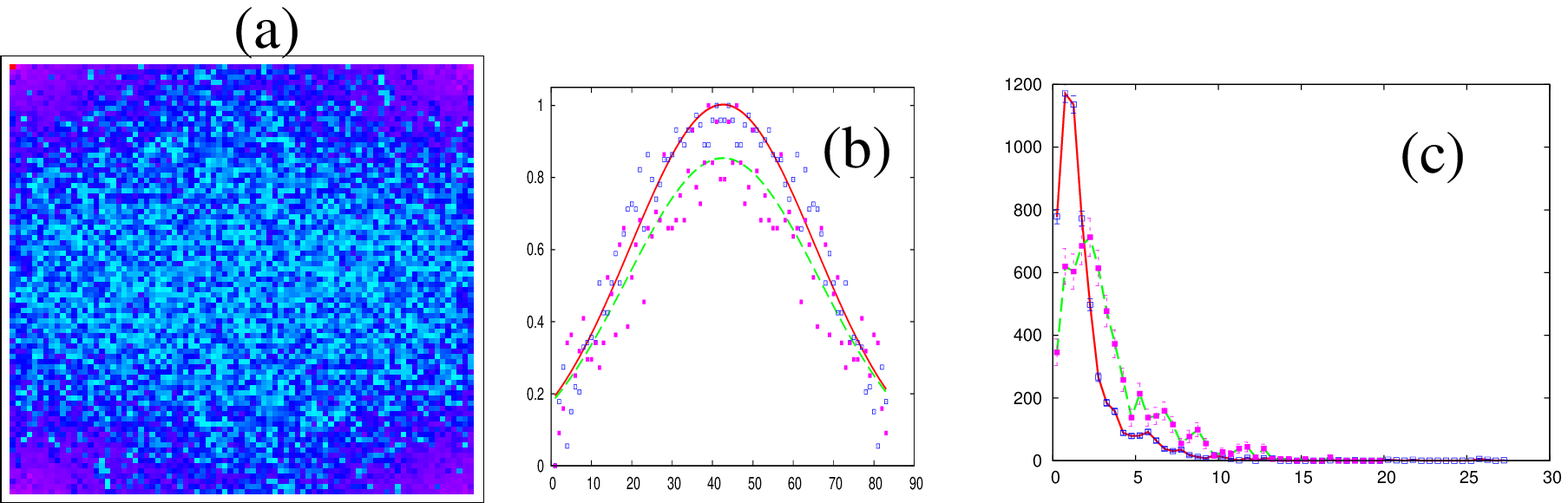}}
\end{center}
\caption{}{For the stretched CMBR data of Fig.5, (a) shows the 
2D Fourier transforms, (b) shows corresponding histograms, and
(c) shows plots of frequency distributions for widths of patches
(excursion sets) of Fig.5 in  X and Y directions for the entire
equitorial belt.}
\label{Fig6}
\end{figure*}

 Even with this limitation, it is important to appreciate that this
simple technique of shape analysis is able to answer an important
question in an almost model independent manner. That is whether the
universe ever expanded anisotropically almost from the beginning of
inflation near $t \simeq 10^{-35}$ sec. all the way up to the stage of last
scattering when the universe was 300,000 years old. We can try to 
put stronger constraint on the anisotropic expansion factor $\alpha$.
 For this, we have repeated the analysis of Fig.6c with different 
values of $\alpha$ (including values of $\alpha < 1$). Figs.7a-d 
show the cases of $\alpha$ varying from 1.5 to 1.2. We see 
that the two histograms corresponding to X and Y widths of patches are 
clearly separated for $\alpha = 1.5$ and 1.35. However, for $\alpha 
\le 1.3$ the differences in the two histograms is insignificant.
It is important to note here that the most important,
qualitative, signature of anisotropic expansion in our technique is the
relative lateral shift of the curves of the X and Y frequency distributions.
This is clearly seen in Fig.6c representing larger length scales
in one direction compared to the other direction. This automatically
is correlated with change in relative heights of the peaks. Thus in Fig.7b
though peak height are very different, our focus on detecting anisotropic
expansion (here with $\alpha = 1.35$) is in the shift of the overall
curve towards right. When we use smaller values of $\alpha$, this
lateral shift is not significant. 
With this, it seems reasonable to conclude that with our analysis technique, 
and with the present CMBR data, one can put a conservative upper 
bound of $\alpha < 1.35$ on the anisotropic expansion in the entire 
history of the universe (which could leave any imprints on the 
superhorizon fluctuations in CMBR, as explained above).

\begin{figure*}[!hpt]
\begin{center}
\leavevmode
\epsfysize=12truecm \vbox{\epsfbox{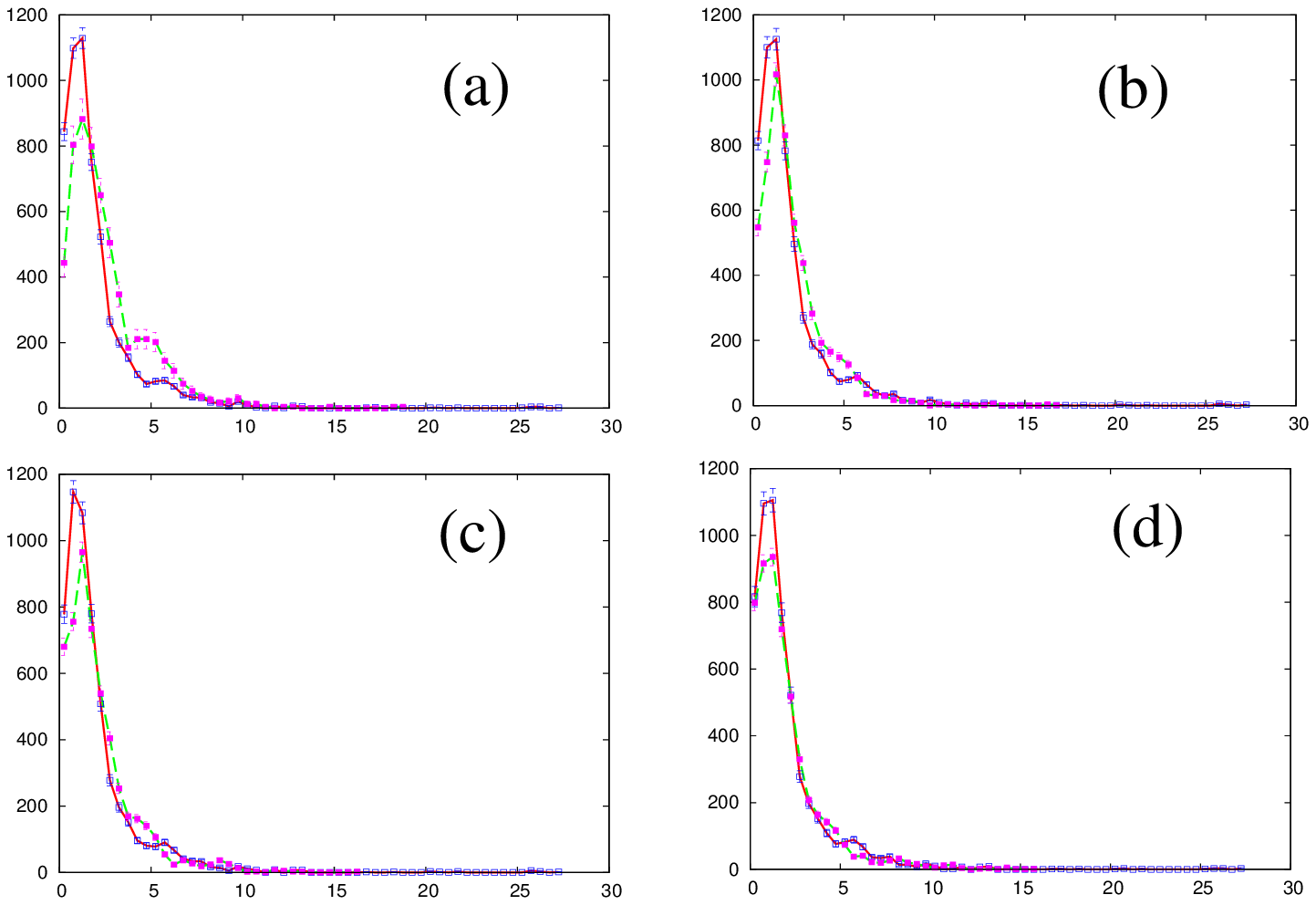}}
\end{center}
\caption{}{Plots of frequency distributions for CMBR data (as in Fig.6c) with
stretch factor $\alpha $ ranging from 1.5 (a), 1.35 (b), 1.3 (c), 1.2 (d).
$X$ and $Y$ width distributions are shown by solid and dashed curves.} 
\label{Fig7}
\end{figure*}

  To rule out any chance coincidence, where any anisotropic expansion
may have taken place along $45^\circ$ from the z axis (which would stretch
many patches by same factor in X and Y directions), we have repeated
the above analysis of the patches along the equatorial belt for various
orientations of the z axis. We have considered the original z axis of 
the WMAP data, and have also rotated the z axis (about the x axis) by
$30^\circ$ and by $70^\circ$ (all the figures have been presented for this
last case of $70^\circ$ rotation, as mentioned above).  The results are 
almost the same for all these cases as shown in Fig.4b, hence we do not 
show these plots. We have also repeated the analysis by using under-density 
patches in CMBR sky and the results are the same.

\section{ANALYSIS WITH SIMULATED FLUCTUATIONS} 

To check the importance of the geometry of these random patches in our 
technique, we have carried out the entire analysis of sections IV and V 
using simulated fluctuation patches with well defined spherical shapes. This 
has the advantage that one can distinguish between various scenarios of
anisotropies, and see whether these different scenarios can be separately
identified in our analysis. For this, we create a 3 dimensional cubic 
lattice in which  overdensities of constant magnitude and specific geometric 
shapes are created at different locations. This represents
a part of the universe enclosing the surface of last scattering. We then 
determine the shapes of the overdensity patches by embedding a surface
of two-sphere $S^2$ (representing the CMBR sky) in this 
cubic region and recording the patches which are 
intersected by this $S^2$. Fluctuation patches on this $S^2$ give us
the simulated CMBR sky. Using this we repeat the above analysis of
sections IV and V and determine histograms of 2D Fourier transforms
and the histograms of widths of filled patches respectively.
For the sake of simplicity of implementation, we do this analysis
for $20^\circ \times 20^\circ$ patch near the equator for both the methods. 
By changing shapes of the fluctuations from spherical 
to ellipsoidal with constant, or varying ellipticity, and/or by making 
their distribution anisotropic, different initial conditions can be 
analyzed. We will discuss four  broad cases in the following.

\subsection{Spherical fluctuations, isotropic distribution}

 First we consider the situation when the fluctuations were generated
by an isotropic process, and the entire evolution of the universe
remains isotropic. In this case initial shapes of the fluctuations,
as well as their distribution will be (statistically)
isotropic, and this feature will be retained throughout the evolution of
the universe leading to final spherical fluctuations. Also,
the distribution of these fluctuations will remain isotropic. This situation
is shown in Fig.8a. We have generated spherical fluctuations of different
sizes. Note that geometrical spherical shape of these
fluctuations is only supposed to represent statistically averaged
shape of (the excursion set of) a realistic fluctuation. Also, note
that all shapes may not appear spherical. This is because we are
analyzing fluctuations on the surface of last scattering which represents
intersection of a 2-sphere with spherical fluctuations present in the 
universe.  Fig. 8b shows the  2D Fourier transform of (a) and (c) shows
the corresponding histograms. Fig.8d shows the histograms of widths of 
intersections of the excursion sets in (a)  with thin X and Y slices 
as discussed in section V. The respective histograms in the two 
directions overlap well showing the isotropy in (a). Note that there are
larger errors in Fig.8d (and similarly for other figures for the
simulated geometrical fluctuations) compared to, say, Fig.4b. This is
because we are only using a $20^\circ \times 20^\circ$ region here
with small number of fluctuation patches, compared to the full equatorial
belt for Fig.4b.  Ratio of the widths of the two Gaussian histograms in 
(c) (for 2D Fourier transform) is about 1.02 $\pm$ .01, again showing 
isotropy. It is clear from results shown in Figs.3,4 for
real CMBR data that if we considered deformed (ellipsoidal) fluctuation
patches with their orientations chosen randomly then again we will
get overlapping histograms. Note that anisotropic expansion of the universe
produces deformed shapes, all aligned in specific direction.

\begin{figure*}[!hpt]
\begin{center}
\leavevmode
\epsfysize=12truecm \vbox{\epsfbox{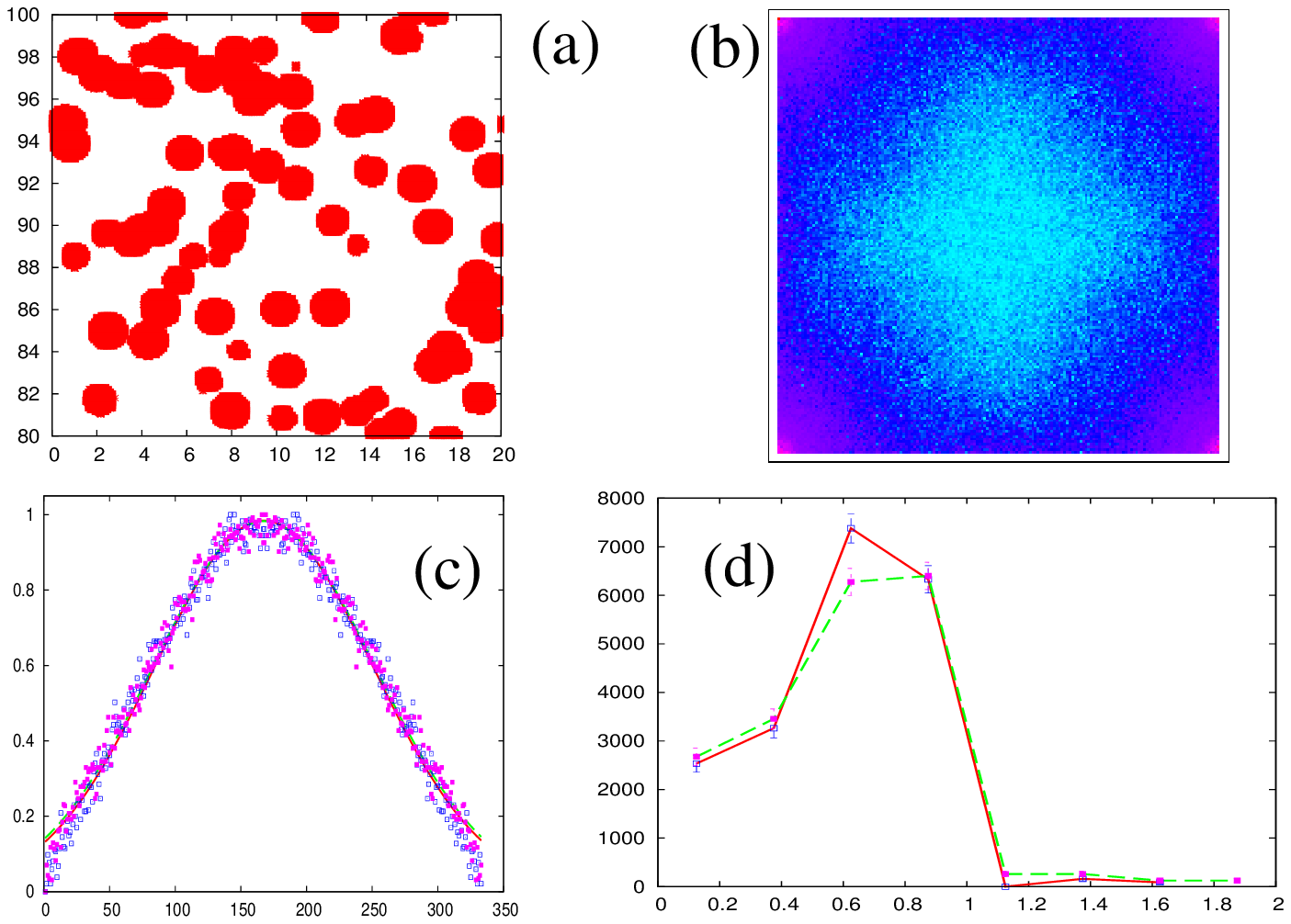}}
\end{center}
\caption{}{This figure shows various plots for the simulated geometrical 
patches.  Again, all the angles are in degrees. (a) shows spherical 
overdensity patches distributed isotropically, (b) shows  
2D Fourier transform (c) shows corresponding histograms, and  
(d) shows the histograms of the widths of the intersections of the 
fluctuation patches with the X and Y spatial slices in (a).}
\label{Fig8}
\end{figure*}

\subsection{Spherical fluctuations, anisotropic distribution}

 It may be possible that the shape of the fluctuations is statistically
spherical however their distribution is not isotropic. Though it is
not clear what detailed model of the universe can give rise to such
a situation, it may be possible, for example, if fluctuations were
generated by topological defects whose production was not isotropic
due to some biasing field. Seeding of topological defects and
density fluctuations arising from them are different physical processes
and may correspond to the present situation. Again we emphasize that
our purpose is not to analyze specific models of density fluctuation
production and their evolutions. We want to analyze different possibilities
of anisotropies arising in the universe. Thus it is immaterial to us
that defect mediated fluctuations are not consistent with CMBR
measurements. Important point to recognize here is that this situation
can only arise when the universe expansion remains isotropic at all
stages. Because any anisotropy in the expansion will always lead to
shape deformations in general. The anisotropic distribution (of
spherical fluctuations), if at all possible, must result from the
production mechanism of fluctuations.
 
 Fig.9a shows spherical fluctuations which are distributed anisotropically.
Fig.9b and 9c show the 2D Fourier transform and corresponding histograms.
(d) shows the histograms of $X$ and $Y$ widths of the excursion sets in (a).
In Fig.9c, the ratio of the widths of the Gaussians (for $k_x$ and $k_y$ 
directions in Fig.9b) is

\begin{equation}
{\sigma_{k_x} \over \sigma_{k_y}} = 1.09 \pm .01
\end{equation}

We note that the 2D histogram method is sensitive
to anisotropic distribution of the patches whereas the spatial width
histograms are insensitive to it. This is as expected because spatial
width method only analyzes the excursion sets and regions between these
sets is not relevant. This is an important difference between the two
techniques as our main purpose is to detect anisotropic expansion, and
as explained above, the present case must correspond to the isotropic
expansion of the universe.
 
\begin{figure*}[!hpt]
\begin{center}
\leavevmode
\epsfysize=12truecm \vbox{\epsfbox{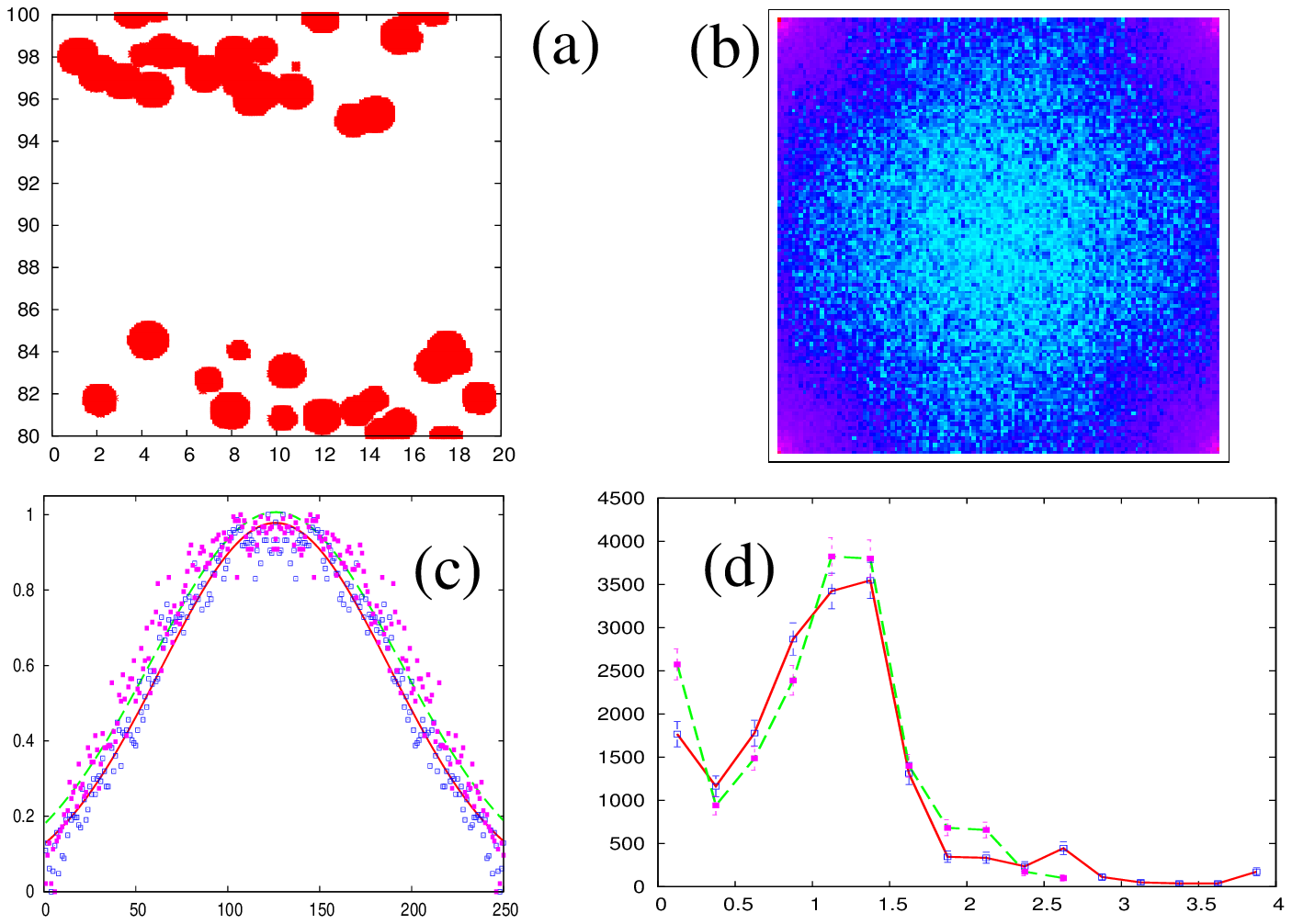}}
\end{center}
\caption{}{(a) shows spherical overdensity patches distributed 
anisotropically. (b) shows 2D Fourier transforms, (c) shows corresponding 
histograms, and  (c) shows the histograms of the widths of
the intersections of the fluctuation patches with the X and Y spatial slices.}
\label{Fig9}
\end{figure*}

\subsection{Ellipsoidal fluctuations, fixed ellipticity}

 This should be the most standard situation when fluctuations are produced
isotropically but they get stretched (and become ellipsoidal) 
during some early transient stage of anisotropic expansion of the
universe. Thus, this is like the case depicted in Figs.5,6, (assuming
inflationary fluctuations at generation were isotropic, but got stretched
later). For this, we introduce the stretching in the simulated data set 
by a factor 2 (as for Fig.5). Fig.10a shows these ellipsoidal fluctuations 
which are distributed anisotropically (with anisotropic stretching the 
distribution automatically becomes anisotropic as in Fig.5). An important 
aspect of these shape deformation is that all these ellipses will have 
fixed ratio of the major axis to minor axis, this ratio directly referring 
to the relative expansion factors of the universe in the two directions.
Fig.10b and 10c show the 2D Fourier transform and corresponding
histograms, and Fig.10d shows the spatial widths histograms for the
excursion sets of (a). Both, Figs.10c,d show distinctly
non-overlapping histograms due to the anisotropy.
The Histograms of 2D Fourier transforms are not fitted by Gaussians here.
To fit these we use a Woods-Saxon potential shaped curve parametrized 
as follows.

\begin{equation}
f(x) = {a \over 1 + exp(abs(x - x_0) - R )/\delta) }
\end{equation}

here $\delta$ gives the width of the region where the function $f(x)$
decays (usually called the skin depth), and $2R$ characterizes the 
central flat part. $x_0$ gives the center. Ratios of $R$ and $\delta$ 
for the two histograms in Fig. 10c are given below.

\begin{equation}
{R_{k_x} \over R_{k_y}} = 1.011 \pm .003 , \qquad \qquad
{\delta_{k_x} \over \delta_{k_y}} = 1.94 \pm .04
\end{equation}

This figure shows that our technique works well for widely
different geometries of patches, from completely random shapes
in Fig.5 to well defined geometries in Fig.10. 
 
\begin{figure*}[!hpt]
\begin{center}
\leavevmode
\epsfysize=12truecm \vbox{\epsfbox{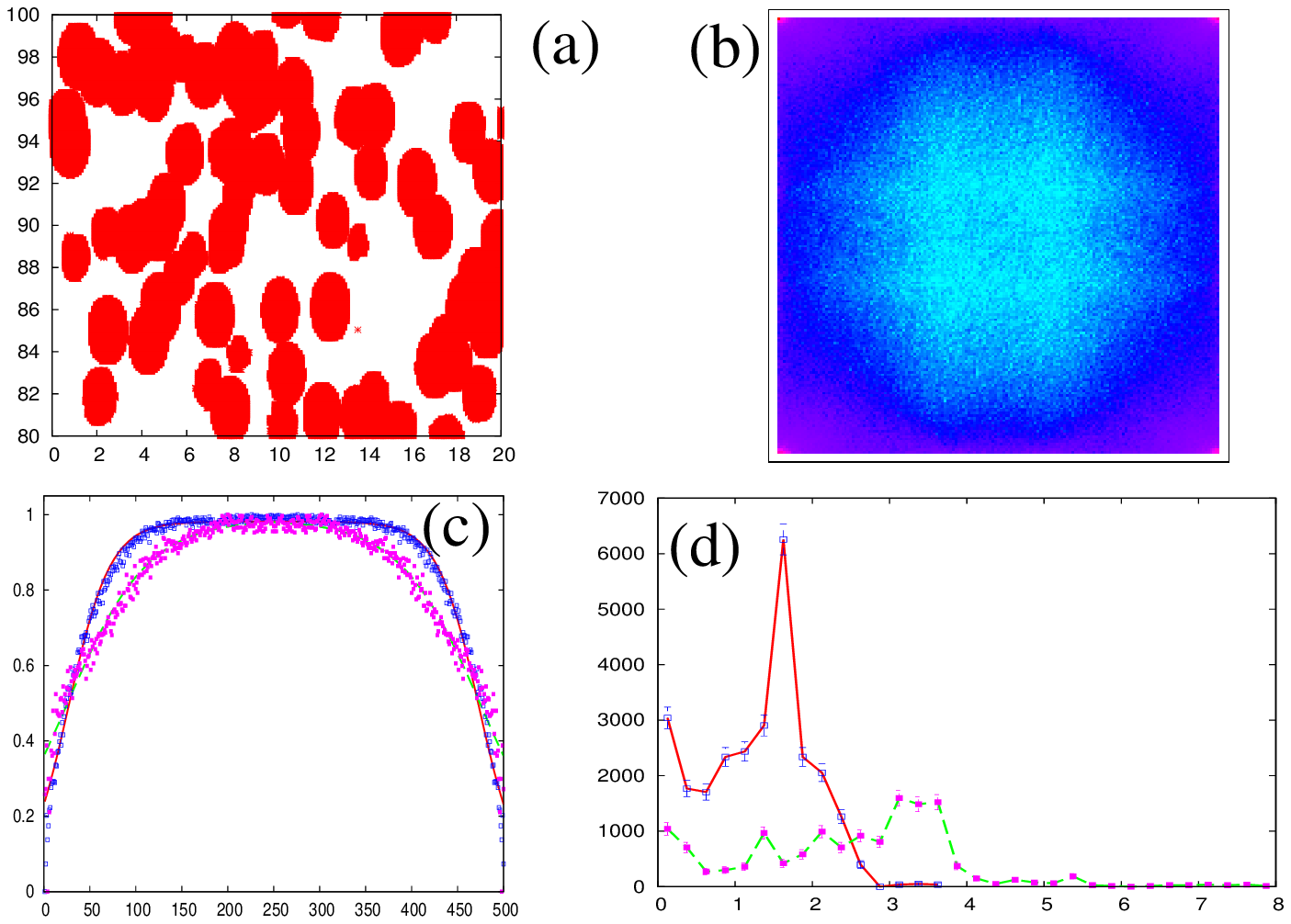}}
\end{center}
\caption{}{(a) shows ellipsoidal overdensity patches distributed 
anisotropically representing anisotropic expansion. (b) shows 2D Fourier 
transforms , (c) shows corresponding histograms, and (d) shows the 
histograms of the spatial widths of the fluctuation patches in (a).}
\label{Fig10}
\end{figure*}

\subsection{Ellipsoidal fluctuations, varying ellipticity}

 We now consider the case when the fluctuations were anisotropic to begin 
with, possibly due to some biasing field, but the universe expansion
remained isotropic throughout. Final fluctuation shapes will clearly
be statistically anisotropic, relating to the initial anisotropy. It
is important to see whether one can distinguish such a situation from
the situation when fluctuations were initially isotropic but
became deformed due to anisotropic expansion of the universe (as shown
in Fig.10). We will argue in the following that it is indeed possible to 
distinguish these two cases with a little bit of physics input about the 
nature of processes generating density fluctuations.
 
  Note that, as argued for Fig.10, when anisotropic shapes of fluctuations
arise from anisotropic expansion, it leads to exactly same ratio of
major to minor axis of the ellipsoids corresponding to the ratio of
expansion factors of the universe in the two directions. This feature
is extremely unlikely in any process of fluctuation production where
anisotropy is arising due to some bias (e.g. some background field).
Typically one will expect certain growth of the anisotropy of the 
fluctuation in the background field implying that the anisotropy factor of 
the fluctuations should vary with the size of fluctuations. Note that 
there is range of sizes expected anyway as fluctuations are produced at 
different times  and hence undergo different expansions. Such a class of 
different sized fluctuations will have same anisotropy factor.
 However, even for fluctuations
produced at any particular given time, there is always a range of
sizes of fluctuations for any physical mechanism, typically relating to
some correlation length. These fluctuations of different initial sizes
will undergo different evolutions in the biasing field (or due to biased
production mechanism itself) and should end up with fluctuations of
different sizes having different ellipticity factors. 
(If the anisotropic expansion factor changes in time then also one
will get this type of situation which may be difficult to distinguish.
However, in that case the pattern of ellipticity variation with size
may be of a different nature.)

 To model this situation we consider ellipsoidal fluctuations with 
different sizes having different ratios for the major to minor axis.
Fig.11a shows distribution of such fluctuation patches. Fig.11b,c show
the 2D Fourier transform and corresponding histograms and Fig.11d shows
the histograms of spatial widths
of the excursion sets. We note that though 2D Fourier transform method
shows difference in the two histograms, thereby detecting anisotropy, the
difference is qualitatively similar to that shown in Fig.10c. 
We again fit these two histograms using the Woods-Saxon function.
Ratios of the relevant parameters (Eq.(3)) for the two histograms in
Fig. 11c are given below.

\begin{equation}
{R_{k_x} \over R_{k_y}} = 0.999 \pm .004 , \qquad \qquad
{\delta_{k_x} \over \delta_{k_y}} = 1.46 \pm .04
\end{equation}

Using this method, thus, it is hard to distinguish the case of Fig.10 
from the present case. (Though, difference in Fig.10b and Fig.11b looks
significant and may be of qualitative nature with a different 
characterization.) On the other hand, Fig.11d shows a qualitative 
difference between the X and Y histograms from that in Fig.10d in terms 
of shift of the peak for Y histogram from the one in X histogram.
The shift of the peak by about factor of 2 in Fig.10 is directly
related to the stretching of each patch by this factor. In contrast,
in Fig.11 the peak reflects the fact that for small patches stretching
is minimal. In that sense, for small scale patches, the structure of
the two curves should be like in Fig.8. The stretching becomes
significant only for large scale patches in Fig.11, consequently,
the two curves differ for large scales. Though we do not know in detail
how to characterize the shapes of the curves in terms of the geometry
and sizes of patches, it is clear that there is a qualitative difference 
in the plots of Fig.11d and Fig.10d. This is very important as it shows 
the possibility that one may be able to distinguish between the 
anisotropic expansion case and anisotropic production process, both 
leading to finally anisotropic fluctuations.

\begin{figure*}[!hpt]
\begin{center}
\leavevmode
\epsfysize=12truecm \vbox{\epsfbox{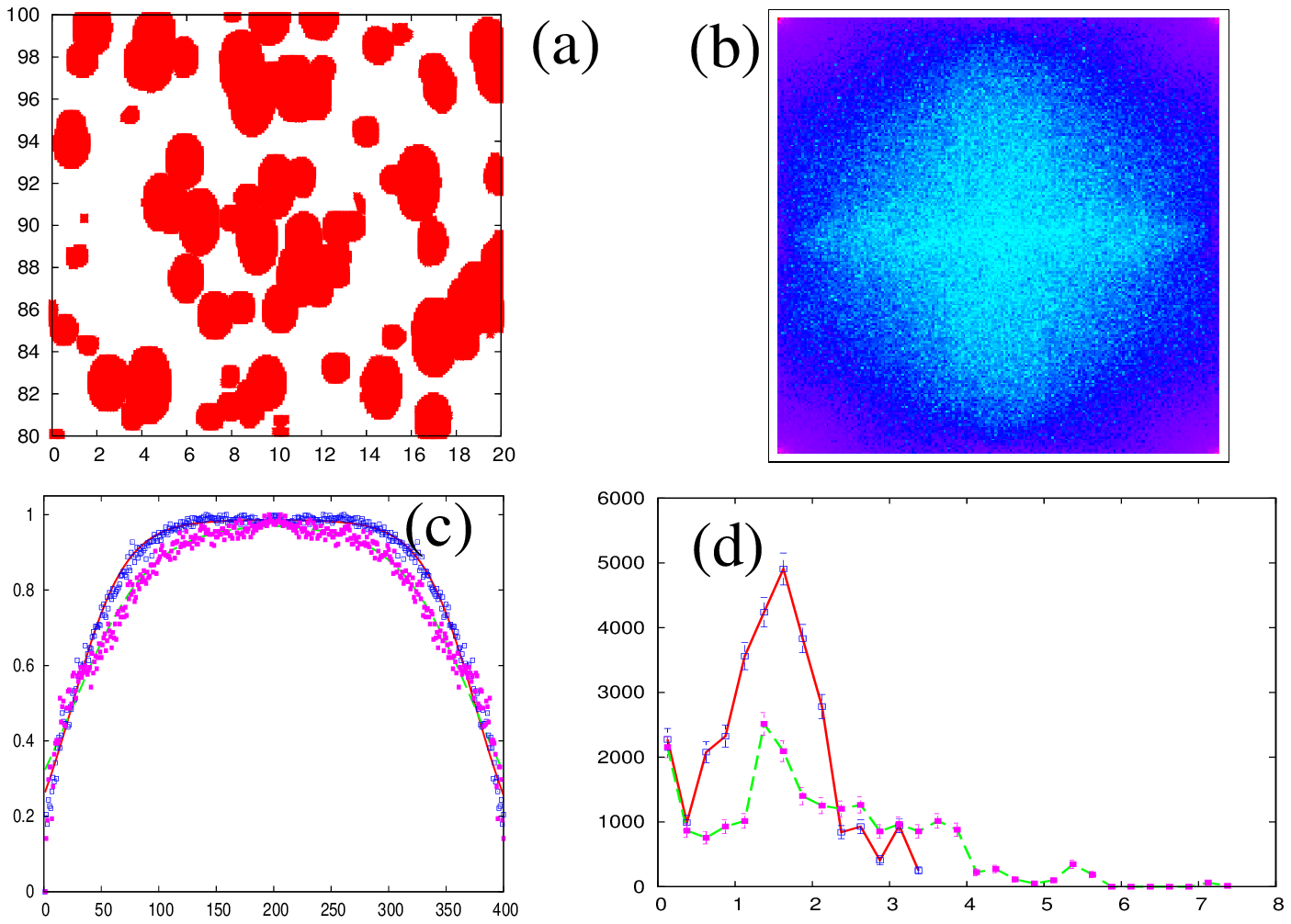}}
\end{center}
\caption{}{(a) shows ellipsoidal overdensity patches with varying
anisotropy factors. (b) shows 2D Fourier transforms, (c) shows
corresponding histogram, and (d) shows histogram of 
the spatial widths of the fluctuation patches in (a).}
\label{Fig11}
\end{figure*}

\section{CONCLUSIONS}

  We emphasize the most important part of our results,
that a simple technique of shape analysis is able to answer an important
question in an almost model independent manner. That is whether the
universe ever expanded anisotropically almost from the beginning of
inflation near $t \simeq 10^{-35}$ sec. all the way up to the stage of last
scattering when the universe was 300,000 years old. Even with our 
qualitative approach of comparing the histograms in the two 
directions, we can conclude that our technique can rule out any
anisotropic expansion of the universe in the past to less than 35\%, 
(apart from any sufficiently early anisotropic stages of inflationary
universe which are followed by very long, isotropic inflationary 
stage). In particular any anisotropic expansion stage after the end
of inflation (or after the end of generation of fluctuations) is 
certainly restricted to $\alpha < 1.35$.
With Planck data one should be able to do shape analysis of
patches with high resolution. It will be specially important to
do the shape analysis of sub-horizon fluctuations as that can provide
additional information about the dynamical details of the evolution 
of subhorizon fluctuations such as any background fields, viscosity etc.

A major drawback of our present analysis is its restriction to relatively
small angular scale (which is tentatively taken to be about $20^\circ$
here). Thus, claims of anisotropy at 
quadrupole level in the literature are not examined by our 
analysis at present. Clearly this is physically very interesting case
to check as any early anisotropic stage of the inflationary universe,
which eventually turns into isotropic expansion, will lead to anisotropic
shapes of patches at large angular scales (those which exited horizon early
during inflation) while patches at small angular scales will be
isotropic in shapes. We are working on improvement of our techniques for
larger angular scales, and also trying to get better control
on statistical fluctuations in our plots.

 We mention here that this technique of identifying anisotropic early
 expansion by shape analysis of superhorizon fluctuation regions has a natural
 application to the physics of relativistic heavy-ion collision experiments
 (RHICE). Measurements of elliptic flow in non-central
 collisions in these experiments has provided a wealth of important
 information about initially anisotropic expansion of the anisotropic
 plasma region, such as equation of state, viscosity etc. Recently, we
 have argued that there are certain similarities in the physics of density
 fluctuations in the universe and in RHICE, in particular, relating to
 the presence of superhorizon fluctuations \cite{cmbhic}. Thus,
 the technique presented in this paper, when suitably
 applied to the shape analysis of large wavelength energy/number density
 fluctuation regions  in RHICE, may  yield important
 information about the early anisotropic expansion of the plasma.
 We will report this work in a future publication. This technique  has 
obvious applications for other systems such as in condensed matter physics,
for example in studies of growth of domains during phase transitions
especially in the presence of biasing fields.

\acknowledgments

  We are very thankful to Tuhin Ghosh and Pramod Samal for help in accessing
the CMBR data. We are very grateful to Raghavan Rangarajan, Tarun Souradeep,
Arjun Berrera, Pankaj Jain, Sanatan Digal, Abhishek Atreya, Anjishnu Sarkar, 
and Ananta Mishra for useful discussions and comments. 
For processing CMBR data we have used the HEALPix package. 
We acknowledge the use of the Legacy Archive for Microwave 
Background Data Analysis (LAMBDA).



\begin{thebibliography}{99}

\bibitem{wald} R.M. Wald, Phys. Rev. {\bf D28} 2118 (1983).

\bibitem{cmbanstrpy} P. Jain and J. P. Ralston, Mod. Phys. Lett. 
{\bf A14} 417 (1999); C. J. Copi, D. Huterer, and G. D. Starkman, 
Phys. Rev. {\bf D70} 043515 (2004); P. K. Samal, R. Saha, P. Jain, and 
J. P. Ralston, Mon. Not. Roy. Astron. Soc. {\bf 396} 511 (2009).

\bibitem{aoe} K. Land and J. Magueijo, Phys. Rev. Lett. {\bf 95} 071301 (2005). 

\bibitem{tplg} A. de Oliveira-Costa, M. Tegmark, M. Zaldarriaga, and 
A. Hamilton,  Phys. Rev. {\bf D69} 063516 (2004). 

\bibitem{mnkwsk} J. Schmalzing, M. Kerscher, and T. Buchert,
In {\it Varenna 1995, Dark matter in the universe} , pp 281-291,
(Proc. of Int. School of Physics Enrico Fermi, Course 82, {\it Dark 
Matter in the Universe}, Varenna, Italy, Jul 25 - Aug 4, 1995,
astro-ph/9508154

\bibitem{mnkcmb} M. Migliaccio et al. Nucl. Phys. {\bf B194}(Proc. Suppl.) 
278 (2009); J. Schmalzing and K.M. Gorski, astro-ph/9710185.
 
\bibitem{mnkstr} J.V. Sheth and V. Sahni. astro-ph/0502105; C. Hikage et al.
(SDSS Collaboration), Publ. Astron. Soc. Jap. {bf 55} 911 (2003).

\bibitem{mnksun} I.S. Knyazeva, N.G. Makarenko, and L.M. Karimova,
Astronomy Rep.  {\bf 54} 747 (2010.

\bibitem{praba} G. Rossi, P. Chingangbam, and C. Park,
arXiv:1003.0272;  P. Chingangbam and C. Park, JCAP 0912 
(2009) 019 

\bibitem{ellps} V.G. Gurzadyan et al. , Mod. Phys. Lett. {\bf A20}
813 (2005).

\bibitem{bianchi} See, Chapt. XI, {\it General relativity - an Einstein 
centenary survey},  edited by, S.W. Hawking and W. Israel, Cambridge 
University Press, Cambridge (1979); G.F.R. Ellis and M.A.H. MacCallum,
Coomun. Math. Phys. {\bf 12} 108 (1969).

\bibitem{bianchicmbr} T. Ghosh, A. Hajian, and 
T. Souradeep, Phys. Rev. {\bf D75} 083007 (2007). 

\bibitem{ansinf} R. V. Buniy, A. Berera, and T. W. Kephart,
Phys. Rev. {\bf D73} 063529 (2006). 

\bibitem{aniscont} T.R. Dulaney and M.I. Gresham, Phys. Rev. {\bf D81} 103532
(2010).

\bibitem{wmap7} Jarosik, N., et.al., 2011, Astrophys. J. Suppl.
{\bf 192} 14 (2011); D. Larson et al., Astrophys. J. Suppl. {\bf 192} 
16 (2011).

\bibitem{ft} D.D. Feo, S.D. Nicola, P. Ferraro, P. Maddalena, and G. 
Pierattini, Pure Appl. Opt. {\bf 7} 1301 (1998); B. Josso, D. R. Burton, 
and M.J. Lalor, Mech. Syst. and Signal Process. {\bf 19} 1152 (2005).

\bibitem{ftfiber} M. Tunak and A. Linka, Fibers and Textiles in
Eastern Europe, {\bf 15}, 64 (2007).

\bibitem{ftmtlrg} R. Holota and S. Nemecek, Applied Electronics, 2002, p:88

\bibitem{cmbhic} A.P. Mishra, R. K. Mohapatra, P. S. Saumia, and
A. M. Srivastava, Phys. Rev. {\bf C 77}, 064902 (2008);
Phys. Rev. {\bf C 81}, 034903 (2010);  R. K. Mohapatra, P. S. Saumia, and
A. M. Srivastava, Mod. Phys. Lett. {\bf A26} 2477 (2011).
\end{thebibliography}
\end{document}